\documentclass[reprint,superscriptaddress,amsmath,amssymb,aps]{revtex4-2}
\usepackage{graphicx, hyperref}
\usepackage{xcolor}
\usepackage{soul}
\begin{document}
\title{Trigonal warping effects on optical properties of
anomalous Hall materials}
\author{Jyesta~M.~Adhidewata}\email{jyesta.ma@gmail.com}
\affiliation{Research Center for Quantum Physics, National Research and Innovation Agency (BRIN), South Tangerang 15314, Indonesia}
\affiliation{Theoretical High Energy Physics Research Division,
Faculty of Mathematics and Natural Sciences, Institut Teknologi
Bandung, Bandung 40132, Indonesia}
\author{Ravanny~W.~M.~Komalig}
\affiliation{Research Center for Quantum Physics, National Research and Innovation Agency (BRIN), South Tangerang 15314, Indonesia}
\affiliation{Advanced Functional Materials Research Group, Faculty of Industrial Technology, Institut Teknologi Bandung, Bandung 40132, Indonesia}
\author{M~Shoufie~Ukhtary}
\affiliation{Research Center for Quantum Physics, National Research and Innovation Agency (BRIN), South Tangerang 15314, Indonesia}
\author{Ahmad~R.~T.~Nugraha}
\affiliation{Research Center for Quantum Physics, National Research and Innovation Agency (BRIN), South Tangerang 15314, Indonesia}
\affiliation{Department of Engineering Physics, School of Electrical Engineering, Telkom University, Bandung 40257, Indonesia}
\affiliation{Research Collaboration Center for Quantum Technology 2.0, Bandung 40132, Indonesia}
\author{Bobby~E.~Gunara}\email{bobby@itb.ac.id}
\affiliation{Theoretical  High Energy Physics Research Division,
Faculty of Mathematics and Natural Sciences, Institut Teknologi
Bandung, Bandung 40132, Indonesia}

\author{Eddwi~H.~Hasdeo}\email{eddwi.hesky.hasdeo@brin.go.id}
\affiliation{Research Center for Quantum Physics, National Research and Innovation Agency (BRIN), South Tangerang 15314, Indonesia}
\affiliation{Department of Physics and Materials Science,
University of Luxembourg, L-1511 Luxembourg, Luxembourg}

\begin{abstract}
The topological nature of topological insulators are related to the symmetries present in the material, for example, quantum spin Hall effect can be observed in topological insulators with time reversal symmetry, while broken time reversal symmetry may give rise to the presence of anomalous quantum Hall effect (AHE). Here we consider the effects of broken rotational symmetry on the Dirac cone of an AHE material by adding trigonal warping terms to the Dirac Hamiltonian. We calculate the linear optical conductivity semi-analytically to show how by breaking the rotational symmetry we can obtain a topologically distinct phase. The addition of trigonal warping terms causes the emergence of additional Dirac cones, which {when combined has a total Chern number of $\mp 1$ instead of $\pm 1/2$}. This results in drastic changes in the anomalous Hall and longitudinal conductivity. The trigonal warping terms also activates the higher order Hall responses which does not exist in a $\mathcal{R}$ symmetric conventional Dirac material. We found the presence of a non-zero second order Hall current even in the absence of Berry curvature dipole. This shift current is also unaffected by the chirality of the Dirac cone, which should lead to a non-zero Hall current in time reversal symmetric systems.
\end{abstract}

\date{\today}
\maketitle

\section{Introduction}


The quantum anomalous Hall effect (QAHE) is the presence of a quantized Hall conductance in a material without an external magnetic field due to time-reversal (TR) symmetry breaking~\cite{Laughlin1981,Haldane1988,qi2011}. Similar to the quantum Hall effect, QAHE is a manifestation of the topological properties of a material~\cite{thouless1982, qi2008,qi2011}. In QAHE, the Hall conductivity is determined by the Chern number, which is a topological invariant given by the integral of the Berry curvature over the Brillouin zone~\cite{Xiao2010,bernevig2013,Yang2012}. As a topological invariant, the Chern number is insensitive to small changes applied to the bulk form of the material. However, several ways have been shown to control the winding number of the wave functions allowing the emergence of higher Chern insulators~\cite{Zhao2020}. Therefore, it is intriguing for us to present a particular case where the Chern number can change due to the breaking of rotational symmetry $\mathcal{R}$. One possible model of QAHE material is a gapped Dirac material with broken TR symmetry in the two-band model~\cite{Hasdeo2019}, which can be used to describe the Chern insulators or magnetic topological insulators~\cite{Tokura2019}.

In this work, we examine the optical conductivity of a broken-TR-symmetry Dirac material with trigonal warping terms. The trigonal warping, as occurs in trigonal or honeycomb lattices, breaks the rotational symmetry of the band structure. We find that the breaking of rotational symmetry {can add new three Dirac cones carrying half integer Chern number with opposite sign from the one at K point giving the total values of Chern number equals  $-1$, in contrast to the regular Dirac dispersion which carries a $+1/2$ Chern number. This opens a new possibility to find higher Chern insulators by tracking the strength of trigonal warping and the location of the Dirac cones.} Another advantage of breaking this symmetry is the activation of higher-harmonic red and longitudinal currents arising as a nonlinear response to external fields since those currents are normally suppressed in the $\mathcal{R}$ symmetric system. It is well-known that a nonlinear anomalous Hall current can be observed in materials without inversion symmetry~\cite{Sodemann2015, Nandy2019}, such as transition metal dichalcogenides~\cite{Taghizadeh2019} and strained graphene~\cite{Battilomo2019} (even in the presence of TR symmetry). This nonlinear current is usually attributed to the presence of a Berry curvature dipole, although similar nonlinear transverse current might arise from other sources, such as the semiclassical Jerk term~\cite{Matsyshyn19}. Here we show that, even with zero Berry curvature dipole, we still have a nonzero {transverse} current contribution from the nonlinear shift current in our Dirac material. The shift current arising from a single Dirac point is unaffected by the chirality of that point, so that it warrants the presence of a {transverse} current even in the TR-symmetric systems.  

We organize this paper as follows. In Sec.~\ref{sec:th}, we describe our model Hamiltonian and the {general band structure of our Hamiltonian. In Sec.~\ref{sec:lincon} we describe the method used for calculation of the linear conductivity and its results. We discuss the effects of the trigonal warping such as a non-trivial Chern number and a van Hove singularity. In Sec.~\ref{sec:nonlincon} we described our calculation of nonlinear conductivity, and we discuss the results of our calculation. Finally, we give conclusion of this work in Sec.~\ref{sec:con}.}

\section{Model and methods}
\label{sec:th}

The Hamiltonian for a two-band material can be written in terms of Pauli matrices $\boldsymbol{\tau} = (\tau_x, \tau_y, \tau_z)$ as follows:
\begin{align}
H = \boldsymbol{\tau} \cdot \mathbf{d},\label{eq:H}
\end{align}
where the matrices work on the vector $\vert \psi \rangle = (\psi_A, \psi_B)$ consisting of the wave function of each site in a lattice. Around the $K$ and $K'$ points of a gapped graphene-like materials , $\mathbf{d}$ is given by:  
\begin{align}
    \mathbf{d} = (\chi vp_x + \lambda (p_y^2 - p_x^2), vp_y + 2 \chi \lambda p_x p_y , \Delta),
    \label{eq:dvect}
\end{align}
 where $\Delta$ is half of the band gap,  $\chi = +1$ ($\chi = -1$) is the chirality around the $K$-point ($K'$-point), and $v$ along with $\lambda$ are constants determined by the material's structure.  Expressed in the tight-binding hopping parameter $t$ and the lattice constant $a$, one can write $v$ and $\lambda$ as $v=\sqrt{3}ta/2\hbar$ and $\lambda=ta^2/8\hbar^2$, respectively. The term quadratic in $p$ generates a trigonal warping of the rotationally-symmetric Dirac energy dispersion, in which the $\lambda$ parameter determines the strength or degree of the warping. 

\begin{figure}[tb]
  \centering\includegraphics[width=85mm,clip]{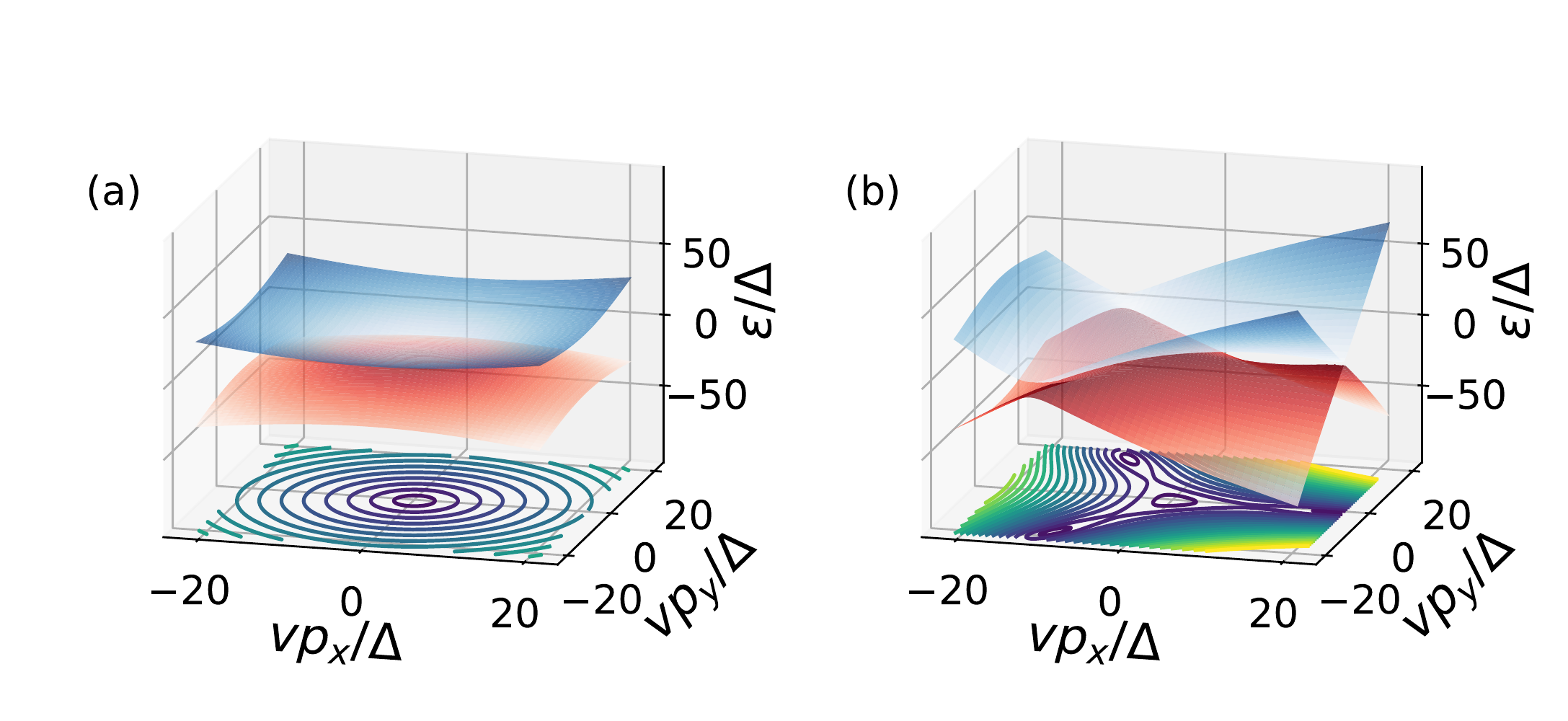}
  \caption{Surface plot of the energy dispersion for the (a) conventional Dirac material and (b) Dirac material with trigonal warping, expressed in the normalized quantities $P = vp/\Delta$ and $d/\Delta$. We take $L=0.05$ for the trigonally warped Dirac material.}
    \label{fig:cont}
\end{figure}

By diagonalizing Eq.~\eqref{eq:H}, we obtain the energy dispersion of our system:
\begin{align}
    \epsilon(\mathbf{p}) =& \pm \vert \mathbf{d} \vert \nonumber\\
    =& \sqrt{\Delta^2 + v^2 p^2 + 2\chi \lambda (3p_y^2 - p_x^2)p_x + \lambda^2 p^4 }.
    \label{eq:dd}
\end{align}
In Fig. \ref{fig:cont}, we plot the energy bands and each corresponding energy contour. In Fig.~\ref{fig:cont}.(a) we plot the energy bands of a conventional Dirac cone ($\lambda=0$) while in Fig.~\ref{fig:cont}.(b) we include the trigonal warping terms with $\lambda \Delta/v^2 = 0.05$. We observe in Fig.\ref{fig:cont}.(b) that with nonzero $\lambda$ the bands no longer form a single symmetric (gapped) Dirac cones, but are warped into a triangular pattern with multiple extrema. From the contour, we can observe the presence of three additional extrema with angular spacing  at $2\pi/3$ around the central point. The distance (in $p$) between the central point and the secondary Dirac points can be found to be $v^2/\lambda \Delta$. 

\section{Linear conductivity}
\label{sec:lincon}

To calculate the linear conductivity of our system, we use the pseudospin approach previously described in \cite{Hasdeo2019}. The dynamics of the electron wavefunction can be described in terms of $\mathbf{m} = \langle \psi \vert \boldsymbol{\tau} \vert \psi \rangle$. The time evolution of $\mathbf{m}$ can be derived from the Bloch equation of motion,
\begin{align}
\frac{d \mathbf{m}}{dt} = \left\langle\psi\bigg\vert \frac{1}{i\hbar}[\boldsymbol{\tau},H]\bigg\vert\psi\right\rangle = \frac{2}{\hbar} \mathbf{d} \times \mathbf{m}.
\end{align}
We can consider an external electric field $\mathbf{E} = E_0 e^{-i\omega t} \mathbf{\hat{x}}$ as a perturbation on $\mathbf{d}$ and expand $\mathbf{d}$ up to the first order as $\mathbf{d}^{(0)} + \delta \mathbf{d}$ where $\delta \mathbf{d} = \frac{i e E_0}{\omega} (\chi v-2\lambda p_x, 2\chi \lambda p_y, 0)$. Then the perturbation in $\mathbf{m}$ is given by
\begin{align}
\frac{d}{dt} \delta \mathbf{m} = -\frac{2}{\hbar} (\delta \mathbf{m} \times \mathbf{d} + \mathbf{m}^{(0)} \times \delta \mathbf{d}),\label{eq:dm}
\end{align}
where $\mathbf{m}^{(0)} = -\frac{\vert \mathbf{d} \vert}{d}$ is the equilibrium value of $\mathbf{m}$.

Let us take $\delta \mathbf{m} = (\delta m_x, \delta m_y, \delta m_z) e^{-i\omega t}$. Thus, we can solve Eq.~\eqref{eq:dm} for $(\delta m_x, \delta m_y, \delta m_z)$ as follows:
\begin{widetext}
\begin{align}
\delta m_x &= \frac{1}{id((\hbar \omega/2)^2 - d^2)} \left(-i(d_y^2+d_z^2)\delta d_x - ((\hbar \omega/2)d_z - i d_x d_y) \delta d_y + ((\hbar \omega/2)d_y + i d_x d_z) \delta d_z \right)\label{eq:dmx}, \\
\delta m_y &= \frac{1}{id((\hbar \omega/2)^2 - d^2)} \left(((\hbar \omega/2)d_z + i d_x d_y )\delta d_x - i(d_x^2 + d_z^2) \delta d_y - ((\hbar \omega/2)d_x - i d_y d_z) \delta d_z \right) \label{eq:dmy},\\
\delta m_z &= \frac{1}{id((\hbar \omega/2)^2 - d^2)} \left(((\hbar \omega/2)d_y - id_x d_z)\delta d_x + ((\hbar \omega/2)d_x + i d_y d_z) \delta d_y - i(d_x^2 + d_y^2) \delta d_z \right)\label{eq:dmz},  
\end{align}
\end{widetext}

The electron current can then be calculated as follows:
\begin{equation}
    \mathbf{j} = \sum_{\mathbf{p}} e \left \langle \frac{\partial H}{\partial \mathbf{p}} \right \rangle = \sum_{\mathbf{p}} e \frac{\partial d_x}{\partial \mathbf{p}} \delta m_x +  \frac{\partial d_y}{\partial \mathbf{p}} \delta m_y.
    \label{eq:jj}
\end{equation}
By substituting Eqs.~\eqref{eq:dmx}) - \eqref{eq:dmz} to Eq.~\eqref{eq:jj} and using $j_i = \sigma_{ij} E_j$, we obtain the Hall conductivity as follows:
\begin{align}
\label{eq:hall0}
\sigma_{xy} = \frac{\chi e^2}{4\pi^2 \hbar^2} \int d^2 p \frac{(\hbar\omega/2) \Delta}{d \omega((\hbar \omega/2)^2 - d^2)}(-4\lambda^2 p^2 + v^2),
\end{align}
where we change the summation on $\mathbf{p}$ to integration and we define $d \equiv \sqrt{\Delta^2 + v^2 p^2 + 2\chi \lambda v (3p_y^2-p_x^2)p_x+\lambda^2 p^4}$.
Furthermore, we define $P \equiv vp/\Delta$ and $L \equiv \lambda \Delta/v$ to normalize the integral and we obtain,
\begin{align}
\sigma_{xy} = \frac{\chi e^2}{8 \pi^2 \hbar} \int d^2 \mathbf{P} \frac{(-4L^2 P^2 + 1)}{D(\Omega^2 -D^2)},
\end{align}
where $D \equiv \sqrt{1+P^2+2\chi L(3P_y^2 - P_x^2)P_x + L^2 P^4}$ and  $\Omega=\hbar\omega/(2\Delta)$. We note that 
\begin{equation}
    \mathcal{F}_{xy} = \frac{\chi \hbar^2 \Delta(-4\lambda^2 p^2 + v_F^2)}{d^{3}},
\end{equation}
is the Berry curvature, such that for $\omega = 0$ we see that the Hall conductivity is given by $e^2 C/h$, where $C = (1/4\pi \hbar^2) \int \mathcal{F}_{xy} d^2 \mathbf{p} $ is the Chern number.

By using similar method, we obtain the longitudinal conductivity $\sigma_{xx}$ as follows:
\begin{widetext}
\begin{align}
\sigma_{xx} &= \frac{-ie^2}{8\pi^2\hbar} \int \frac{(1+P_y^2+4L^2 P^2 + 4 L^2 P_y^2 (4 P_x^2 + P_y^2) + 4L^4 P_y^2 P^4) - 4\chi L P_x(1 + P_y^2- 2 L^2 P_y^2 P^2)}{D \Omega (\Omega^2 - D^2)} d^2 P. \label{eq:sxx}
\end{align}
\end{widetext}
It is noted that due to the symmetry of the band structure under a $120^\circ$ rotation, the longitudinal conductivity is isotropic, in particular $\sigma_{xx} = \sigma_{yy}$. The integrands in the $\sigma_{xx}$ and $\sigma_{yx}$ have a singularity at $D = \pm \Omega$ whenever $ \vert \hbar \omega \vert \geq 2\Delta$. To remove the singularity during the calculation of conductivity we replace $\Omega$ to $\Omega - i\eta$ and taking the limit $\eta \rightarrow 0$. Then, by using the Sokhostski-Plemelj relation, we obtain the imaginary (real) part of $\sigma_{xy}(\Omega)$ ($\sigma_{xx}$ and $\sigma_{yy}$)
. To obtain the real (imaginary) part of $\sigma_{xy}$ ($\sigma_{xx}$ and $\sigma_{yy}$) we can use the Kramers-Kronig relations
\begin{align}
\textrm{Re}[\sigma_{xy}(\Omega)] &= \frac{2}{\pi} \int_0^{\infty} \frac{\Omega' \textrm{Im}[\sigma_{xy}(\Omega')]}{\Omega'^2 - \Omega^2} d\Omega'\\
\textrm{Im}[\sigma_{xx}(\Omega)] &= -\frac{2}{\pi} \int_0^{\infty} \frac{\Omega \textrm{Re}[\sigma_{xx}(\Omega')]}{\Omega'^2 - \Omega^2} d\Omega' \\
\end{align}
The delta function integrals, and the Kramers-Kronig integrals, can then be calculated numerically. We used Python's SciPy library~\cite{SciPy2020} for the numerical integrations. 

\begin{figure}[tb]
  \centering\includegraphics[width=85mm,clip]{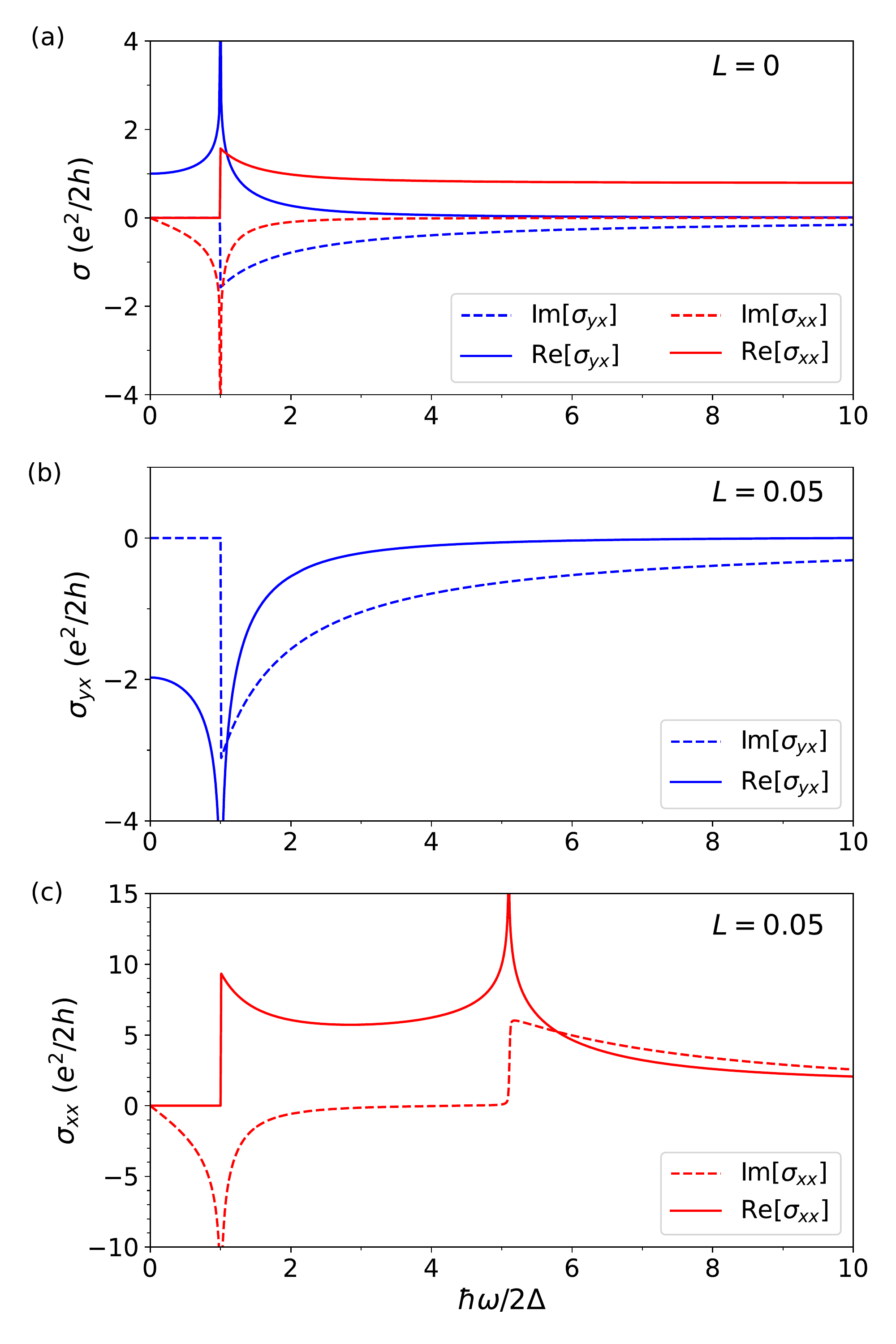}
  \caption{(a) The linear optical conductivity for conventional Dirac material with $L = 0$.  (b) The Hall conductivity and (c) The longitudinal conductivity for Dirac material with trigonal warping ($L = 0.05$). The $x$ axis shows the normalized frequency $\Omega = \hbar \omega/2\Delta$.}
    \label{fig:lin1}
\end{figure}

For our numerical calculation, we set $L = \Delta/6t$ to $L = 0.05$, and assume the bandgap $\Delta$ to be nonzero. We plot the linear conductivity of our model as a function of normalized frequency $\Omega = \hbar \omega/2 \Delta$ in Fig. \ref{fig:lin1}. For comparison, we plot the conductivity of a conventional Dirac material with $L=0$ in Fig.~\ref{fig:lin1}(a) and the conductivity of a Dirac material with trigonal warping material in Figs.~\ref{fig:lin1}(b)--(c). We take the warping coefficient $L=0.05$, which roughly corresponds to a band gap of $\Delta = 0.5~\mathrm{eV}$ and a first neighbour hopping integral similar to graphene, $t = 2.76~\mathrm{eV}$. We note from Eq.~\eqref{eq:hall0} that in case of zero bandgap, the Hall conductivity will vanish.    
From these results, we see that the DC Hall conductivity for our model is twice higher than the DC Hall conductivity for a Dirac material, meaning the Chern number of our trigonal warping band structure is $C=\pm 1$, in contrast to the conventional $C=\pm1/2$ Chern number for a single Dirac cone. This is due to the presence of three extra Dirac cone around the central Dirac point created by the trigonal warping term (see Fig. \ref{fig:cont}). By integrating the Berry curvature around each of these Dirac cones, it can be seen that the central cone carries a $C=\pm1/2$ Chern number while the secondary cones carry a $ C'=\mp1/2$ Chern number, yielding a total of $\mp1$ overall Chern number. This presence of additional Dirac points is reminiscent of the model examined in Ref.\cite{Sticlet2013} concerning a Haldane-model like material with third and higher-neighbor hoppings. However, we must note that here we only consider a single Dirac point, while generally in materials with intact TR symmetry, each Dirac cone is paired with another Dirac point with opposite chirality and opposite Chern number. As a result, when accounting for all the Dirac cones in the material, the total Chern number is zero, unless we break its TR symmetry in which we can obtain a non-zero Chern number. {Still, our result demonstrates that trigonal warping allows the presence of a nontrivial Chern number under the appropriate parameters. For example, in the third neighbor hopping model of \cite{Sticlet2013} we have three additional Dirac cone in a trigonal configuration, which give a total Chern number of $\pm 2$ for a certain range of the third-neighbor hopping, while} with $\mathrm{LaN_2}$, there are six Dirac cones in the Brillouin zone, which results in a total Chern number of $C=3$~\cite{Li20}. 

Another notable feature from Fig. \ref{fig:lin1} is the presence of a singularity at $\Omega \approx 5$ for the longitudinal conductivity, in addition to the singularity at resonant frequency $\hbar \omega = 2\Delta$. This singularity is related to the van Hove singularity, where the density of states diverge due to a saddle point in the band structure. By taking the derivative of Eq.~\eqref{eq:dd}, we can verify that a saddle point of $\vert \mathbf{d}(p) \vert$ occurs when $\vert \mathbf{d} \vert = \sqrt{1+1/(16L^2)} = 5.09$ for $L = 0.05$. This saddle point agrees with previous calculations of optical conductivity  \cite{Hipolito2016,Pratama2020}, which shows the presence of a singularity for materials with honeycomb lattice. Additional results regarding Kerr and Faraday angles related to this material are given in the Appendix.    

\section{Nonlinear conductivity}
\label{sec:nonlincon}
To further explore the consequences of our band structure's asymmetry, we next calculate the nonlinear dc response of the trigonally warped Dirac material. Following the method given by \cite{Taghizadeh2019}, we can write the current as the sum of intraband $\mathbf{j}_\textrm{intra}$ and interband $\mathbf{j}_\textrm{inter}$ contributions. We first note that the intraband contribution from the anomalous photocurrent term~\cite{Rostami2018} is zero for our system. For a 2D system with normally-incident light, the anomalous photocurrent term is reduced to the curl of the Berry curvature $\mathbf{\mathcal{F}}_{xy} = (\partial_{k_x} \mathcal{A}_y - \partial_{k_y} \mathcal{A}_x) \mathbf{\hat{z}}$,
\begin{equation}
    \mathbf{j}^{(2)}_{ap}(0) = -\frac{ie^3}{2h^2 \omega} \int_{-\infty}^{\infty} \int_{-\infty}^{\infty} d^2 \mathbf{k} (\mathbf{E}\times \mathbf{E}^*) \times (\nabla \times \mathbf{\mathcal{F}}_{xy}). 
\end{equation}
Expanding the integrand into component form, we see that the anomalous photocurrent can be expressed in terms of the Berry curvature dipole (see \cite{Sodemann2015}) $D_i$ as follows:
\begin{equation}
    j^{(2)}_{\mathrm{ap},i}(0) = -\frac{ie^3 (E_x E_y^* - E_x^* E_y) D_i}{2\hbar^2 \omega},
\end{equation}
where
\begin{equation}
    D_i = \int d^2 \mathbf{k}~\partial_i \mathcal{F}_{xy}.
\end{equation}
The Berry curvature for our system is
\begin{equation}
    \mathcal{F}_{xy} = \frac{\chi \hbar^2 \Delta(-4\lambda^2 p^2 + v^2)}{(\Delta^2 + v^2 p^2 - 2\chi \lambda v p^3 \cos{3\theta} + \lambda^2 p^4)^{3/2}},
\end{equation}
which has three mirror symmetry planes at $\theta = 0^{\circ}, 120^{\circ},$ and $240^{\circ}$. As the derivative of the Berry curvature behaves as a pseudovector under reflection symmetry, the Berry curvature dipole will always be zero, guaranteeing zero {photocurrent} unless this symmetry is broken (such as by an uniaxial strain as in \cite{Battilomo2019}). However, unlike the Berry curvature, the Berry connections $\mathcal{A}_{cc}, \mathcal{A}_{vv}$ does not possess this threefold symmetry. This lack of symmetry allows the presence of a nonlinear {transverse} current arising from the interband shift current. {Therefore}, here we are most interested in the second-order contribution from the shift current,
\begin{equation}
    \mathbf{j}_\textrm{inter}^{(2)}=-\frac{e^2}{\hbar} \sum_{n\neq m,\mathbf{k}} \rho_{nm}\mathbf{ \mathcal{D}_{mn}}(\mathbf{E}(t) \cdot \mathbf{\mathcal{A}_{mn}}).
\end{equation}
Here, $\rho_{nm}$ is the density matrix element, $\mathbf{\mathcal{A}_{mn}} = i\langle \psi_{m} \vert \frac{\partial}{\partial \mathbf{k}} \vert \psi_{n} \rangle$ is the Berry connection vector, and $\mathbf{\mathcal{D}_{mn}} = \frac{\partial}{\partial \mathbf{k}} + i(\mathbf{\mathcal{A}_{mm}} - \mathbf{\mathcal{A}_{nn}})$. 

The second-order contribution of the shift current term can be written in term of the shift vector $\mathbf{r}_{\textrm{shift}} = \mathbf{\mathcal{A}}_{cc} - \mathbf{\mathcal{A}}_{vv} - \partial_{\mathbf{k}} \mathrm{arg}[\mathbf{E} \cdot \mathbf{\mathcal{A}}_{vc}]$\cite{Sipe2000, Sinitsyn2006, Ahn2020, Shi2021},
\begin{equation}
    \mathbf{j}^{(2)}_\textrm{\textrm{shift}} = \frac{e^3}{2h} \int d^2 \mathbf{k} \delta(\hbar \omega - 2d) \vert \mathbf{E} \cdot \mathbf{\mathcal{A}}_{vc}^* \vert^2 \mathbf{r}_\textrm{shift}.
    \label{eq:shift}
\end{equation}
For linear electric fields in the $x$ or $y$ direction, the conductivities are then
\begin{equation}
\sigma_{ijj}(\omega) = \frac{e^3}{2h} \int d^2 \mathbf{k}~ \delta(\hbar \omega -2d) \mathcal{A}_{vc,j}^* \mathcal{A}_{vc,j} r_\textrm{shift,i}
\label{eq:shiftcond}
\end{equation} 
As the shift vector $\mathbf{r}_{shift}$ is always real, this equation shows the real part of $\sigma_{ijj}$. The imaginary parts of $\sigma_{ijj}$ can be calculated from the Kramers-Kronig relations. 

\begin{figure}[tb]
  \centering\includegraphics[width=85mm,clip]{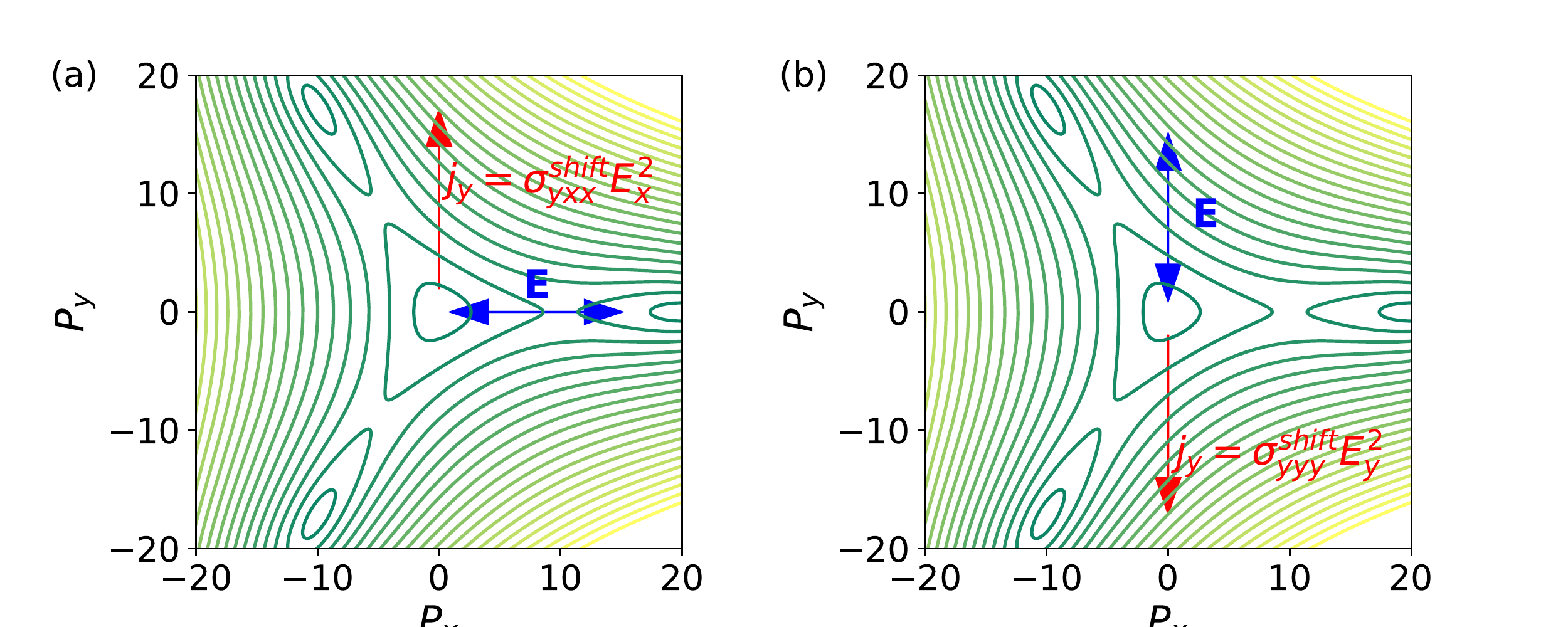}
  \caption{The direction of the current $\mathbf{j}^\textrm{shift}$ and the electric field $\mathbf{E}$ polarization superimposed on the energy contour plot for (a) $x$-polarized electric field (b) $y$-polarized electric field. }
    \label{fig:illshift}
\end{figure}

We now consider the dc current generated by a linearly polarized light with normal incident. The resulting shift current is directed to the $+y$ direction when $\mathbf{E}$ is polarized in the $x$ direction, while if $\mathbf{E}$ is polarized in the $y$ direction the resulting shift current is directed to the $-y$ direction (Fig.~\ref{fig:illshift}).  
This anisotropy in current direction is related to the underlying symmetry of our Hamiltonian. The energy dispersion of our system is invariant with respect to reflection over the $p_x$ axis ($d(p_x,p_y) = d(p_x,-p_y)$) while the Hamiltonian itself obeys $H(p_x,-p_y) = H^*(p_x,p_y)$ as can be seen by simple substitution on Eq.~\eqref{eq:dvect}. Then, this symmetry imposes the following condition on the eigenvalues $\vert u_m (k_x, k_y) \rangle$
\begin{align}
    \vert u_m (k_x, -k_y) \rangle &= \vert u_m (k_x, k_y) \rangle^*,
\end{align}
where $m = c$ for the conduction band and $m = v$ for the valence band. The Berry connections then obey the following conditions,
\begin{align}
    (\mathcal{A}_{mn})_x (k_x, -k_y) &= -(\mathcal{A}_{mn})^*_x (k_x, k_y), \\
    (\mathcal{A}_{mn})_y(k_x, -k_y) &= (\mathcal{A}_{mn})^*_y (k_x, k_y).
\end{align}
As the diagonal Berry connections $\mathbf{\mathcal{A}}_{mm}$ are real, 
\begin{align}
    (\mathcal{A}_{mm})_x (k_x, -k_y) &= -(\mathcal{A}_{mm})_x (k_x, k_y), \\
    (\mathcal{A}_{mm})_y(k_x, -k_y) &= (\mathcal{A}_{mm})_y (k_x, k_y),
\end{align}
while $\vert \mathcal{A}_{vc}(k_x,-k_y) \vert = \vert \mathcal{A}_{vc}(k_x,k_y)\vert$ and
\begin{align}
    \partial_{k_x} \mathrm{arg}[\mathbf{E}\cdot \mathcal{A}_{vc}](k_x, -k_y) &= -\partial_{k_x} \mathrm{arg}[\mathbf{E}\cdot \mathcal{A}_{vc}](k_x, k_y), \\
    \partial_{k_y} \mathrm{arg}[\mathbf{E}\cdot \mathcal{A}_{vc}](k_x, -k_y) &= \partial_{k_y} \mathrm{arg}[\mathbf{E}\cdot \mathcal{A}_{vc}](k_x, k_y).
\end{align}
Thus we can conclude that $(\mathbf{r}_\textrm{shift}(k_x,-k_y))_x = -(\mathbf{r}_\textrm{shift}(k_x,k_y))_x$ while {$(\mathbf{r}_\textrm{shift}(k_x,-k_y))_y = (\mathbf{r}_\textrm{shift}(k_x,k_y))_y$}. As a result, with integration over $k_y$, the $x$ component of the shift current vanishes. Note that the addition of the trigonal warping terms to the Dirac Hamiltonian breaks a similar symmetry with respect to $k_x \rightarrow -k_x$, allowing a nonzero shift current in the $y$ direction.  

\begin{figure}[tb]
  \centering\includegraphics[width=85mm,clip]{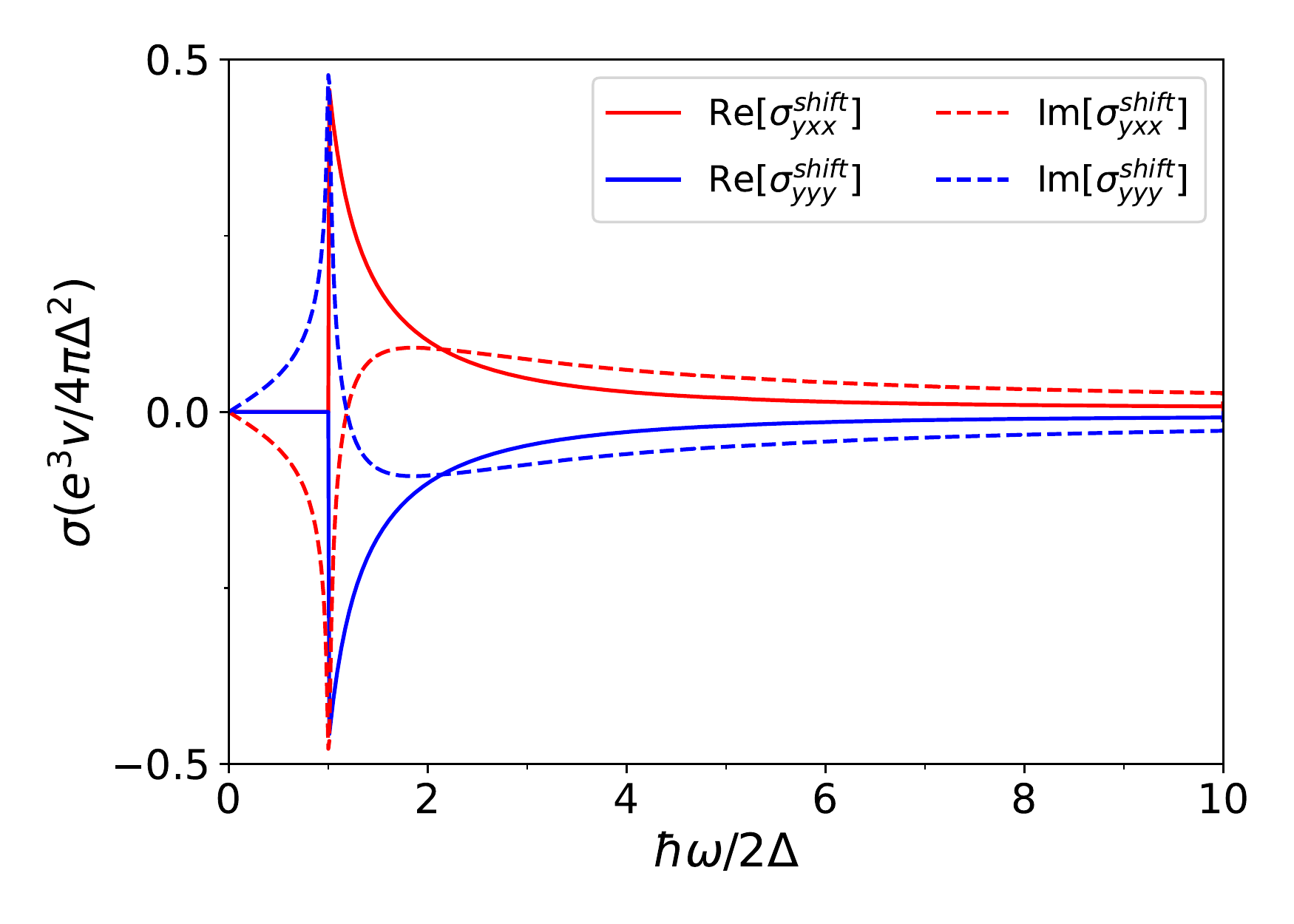}
  \caption{The nonlinear transverse conductivity (in dimensionless units) as a function of normalized light frequency $\hbar \omega/2\Delta$. We plot the second order contribution from the shift current, stimulated by linearly polarized light in the $x$ direction (red) and in the $y$ direction (blue) }
    \label{fig:nonlin}
\end{figure}

We plot the shift conductivity in Fig. \ref{fig:nonlin}. There are several notable qualities from this plot. First, we see that $\sigma_{yxx} = -\sigma_{yyy}$, as required by the $C_3$ symmetry of the system \cite{Hipolito2016}. Also, since the shift current is created by interband transitions only, we see that for frequencies below the band gap the real part of the conductivity is zero, both for the longitudinal ($\sigma_{yyy}$) and transversal ($\sigma_{yxx}$) current. A singular peak is observed when $\hbar \omega = 2\Delta$, as in the linear case. We can also check the effects of different chirality $\chi = \pm 1$ on the nonlinear conductivity. If we consider the remaining nonzero shift conductivity, $\sigma_{yyy}^\textrm{shift} = -\sigma_{yxx}^\textrm{shift}$, all the Berry connections on the integrand, $\mathcal{A}_{ij,y} = i\langle \psi_i \vert \partial_{k_y} \vert \psi_j \rangle$, depend on $\chi$ only through $\vert \psi_i \rangle$, which can be written in terms of the  two-band Hamiltonian $\mathbf{d}$ vector as \cite{bernevig2013}
\begin{align}
\vert \psi_{v} \rangle &= \frac{1}{\sqrt{2d(d+d_z)}} \begin{bmatrix} d_z+d \\ d_x - id_y \end{bmatrix} \\
\vert \psi_{c} \rangle &= \frac{1}{\sqrt{2d(d-d_z)}} \begin{bmatrix} d_z-d \\ d_x - id_y \end{bmatrix}
\end{align}
We see from the Hamiltonian that the chirality $\chi$ is always coupled with $p_x$, thus flipping the chirality is equivalent to a substitution $p_x \rightarrow -p_x$ to the Hamiltonian. So, we have $\mathcal{A}_{ij,y}^{\chi = -1}(k_x, k_y) = \mathcal{A}_{ij,y}^{\chi =+1}(-k_x, k_y)$. As flipping the chirality is equivalent to the substitution $p_x \rightarrow -p_x$, while Eq. (\ref{eq:shift}) is already integrated over $p_x \in (-\infty,\infty)$, we have $\sigma_{ijj}^{\textrm{shift},\chi =+1} = \sigma_{ijj}^{\textrm{shift},\chi =-1}$, which has also been checked manually through numerical integration. As a result, this shift current is nonzero even with TR symmetry, in contrast to the linear Hall conductivity~\cite{Watanabe21}, which vanishes when summed over $\chi=+1,-1$ unless TR symmetry is broken.   

\begin{figure}[tb]
  \centering\includegraphics[width=85mm,clip]{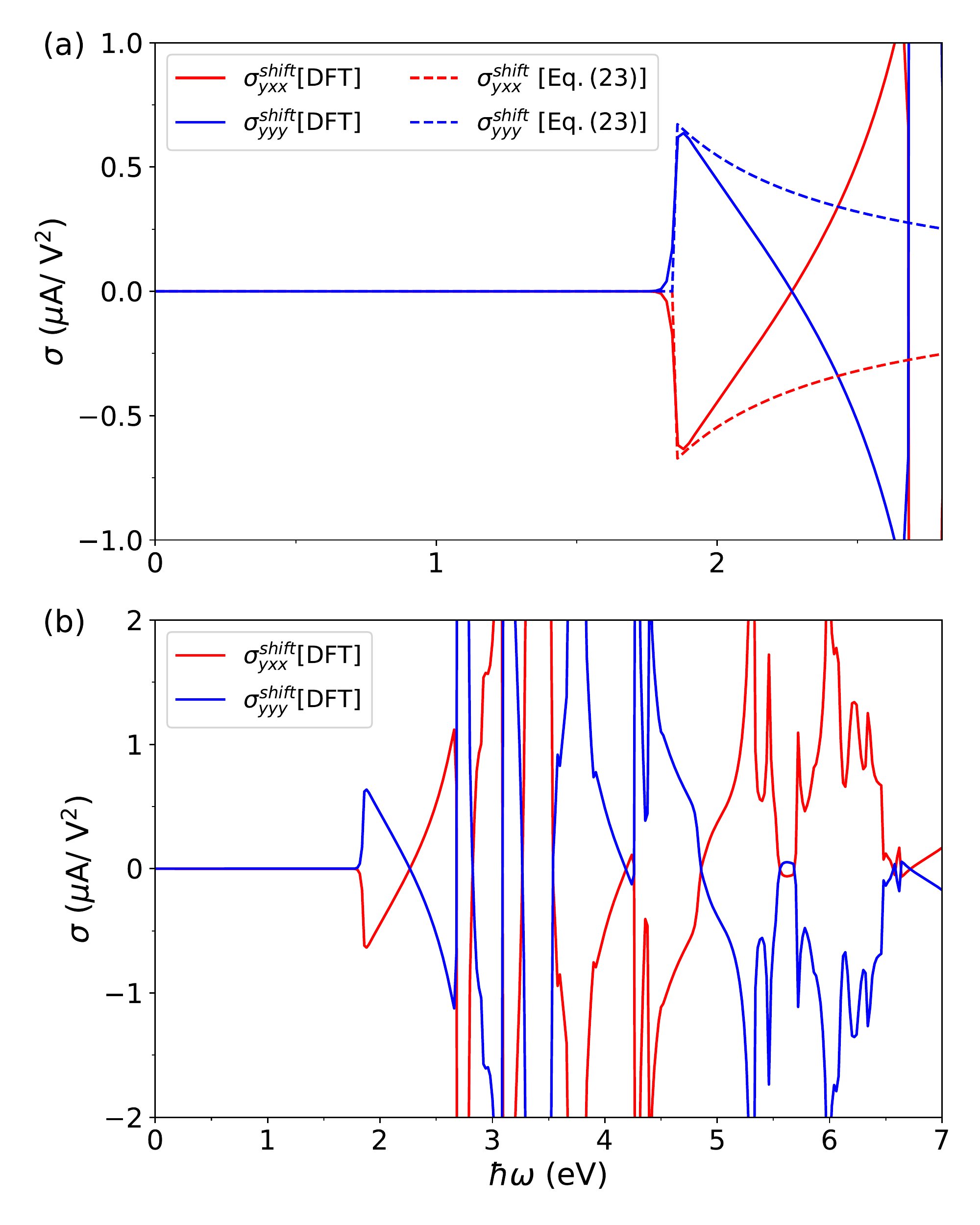}
  \caption{The shift current in $\mathrm{MoS_2}$ as a function of frequency. (a) We compare the results of the first-principles approach (solid lines) with our theoretical approach (dotted lines) at frequencies near the band gap, where interband effects are negligible (b) The first principle calculation of the shift current for the full frequency range.}
    \label{fig:MoS2shift}
\end{figure}

To discover possible realizations for our system, we calculated the shift current of a monolayer $\mathrm{MoS}_2$. $\mathrm{MoS}_2$ possesses a hexagonal lattice, and it has been noted that the band structure around the K and K' points exhibits trigonal warping~\cite{Kormanyos13}. By fitting our Hamiltonian (Eq.~\eqref{eq:H}) to the band structure of $\mathrm{MoS}_2$ around the K point, we obtain the following value of the band parameters: $\Delta  = 0.89~\mathrm{eV}$, $v=0.22~\mathrm{nm/fs}$ and $L = \lambda \Delta /v^2 = 0.065$. We then calculate the shift current of $\mathrm{MoS}_2$ using these parameters and Eq.~\eqref{eq:shift} and also from a first principle approach using Wannier90. We plot the calculated shift current of $\mathrm{MoS}_2$ in Fig.\ref{fig:MoS2shift}. In Fig.\ref{fig:MoS2shift}.(a) we compare our theoretical result from Eq.\eqref{eq:shiftcond} with the DFT results, at frequencies around the band gap, where the effects of multiband interactions are negligible, and in Fig.\ref{fig:MoS2shift}.(b) we show the DFT results for the full frequency range. We note that as the DFT software calculates the 3D conductivity, to fit our theoretical results with the DFT calculations' units we divide Eq.\eqref{eq:shiftcond} with the confinement length, which we take to be 12 nm. The first principle results diverges from the theoretical results for frequencies further from the band gap, with the shift current taking negative values for $\hbar \omega > 2.3~\mathrm{eV}$. However, we emphasize that in our calculations we assumed a two band model, while we believe this divergence to be a result of multiband interaction, which are activated at higher frequencies. We note that Eq.~\eqref{eq:shift} does allow a negative value of the shift current as $r_{{\rm shift},i}$ might take negative values, depending on $\mathcal{A}_{cc} - \mathcal{A}_{vv}$.

\section{Conclusions}
\label{sec:con}

We have shown that around a Dirac point with trigonal warping, the appearance of three additional Dirac valleys with Chern number opposite to the central Dirac point {unlocks the possibility of an anomalous Hall material with higher Chern number from multi-valley contributions.}. Furthermore, by breaking rotational symmetry we show the existence of a nonlinear transverse current with a zero Berry curvature dipole even in the absence of broken TR symmetry. These results provide a new perspective and directions for exploring topological materials.

\begin{acknowledgements}
  We acknowledge HPC BRIN for their high-performance computing facilities.  
  J.M.A. and B.E.G. are supported by  P2MI FMIPA ITB and P2MI KK FTETi ITB.  E.H.H. acknowledges supports from the National Research Fund Luxembourg under grants CORE C20/MS/14764976/TOPREL and C21/MS/15752388/NAVSQM.
\end{acknowledgements}

\appendix

\section{Kerr and Faraday angles}

\begin{figure}[tb]
  \centering\includegraphics[width=85mm,clip]{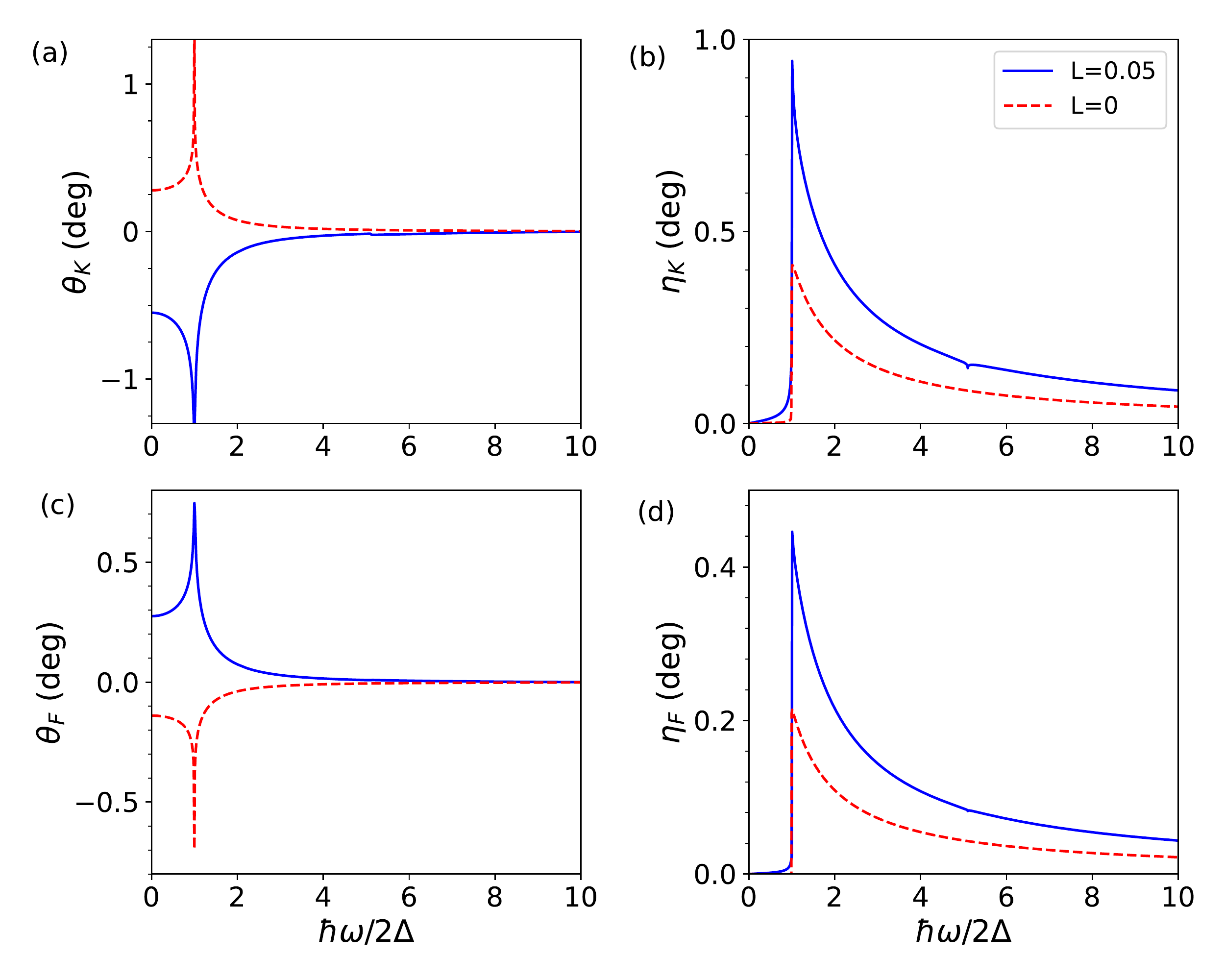}
  \caption{(a) Kerr angle, (b) Kerr ellipticity, (c) Faraday angle, and (d) Faraday ellipticity of a linearly polarized light that is incident in the normal direction to the material.  All properties are plotted as a function of normalized frequency. We take the refractive index of the substrate to be $n=2$.}
    \label{fig:kerr}
\end{figure}

In addition to our main result above, to illuminate the possible methods for the characterization of warping effects, we also plot the resulting Kerr and Faraday angles produced by a linearly polarized light normally incident on a thin film of our anomalous Hall material. For a 2D material on a substrate with refraction index $n$ expression for Kerr angle and ellipticity is given by~\cite{Pratama2020}
\begin{align}
    \tan{2\theta_{K}} &\sim 2 \mathrm{Re}\left[\frac{8\pi \sigma_{yx}}{c(n^2-1)}\right] \\
    \sin{2\eta_{K}} &\sim 2 \mathrm{Im}\left[\frac{8\pi \sigma_{yx}}{c(n^2-1)}\right]
\end{align}
while the Faraday angles is approximately one half of the Kerr angles. We plot the results in Fig.~\ref{fig:kerr}. 

\section{First-principles calculations}

We carried out supporting first-principles, density functional theory (DFT) calculations using the Quantum ESPRESSO  package~\cite{qe2009}. The monolayer 2H-MoS$_{2}$ lattice parameters are adopted from previous calculations \cite{schankler21} with vacuum space set to 20 \AA~in $z$-direction. We utilized generalized gradient approximation exchange-correlation functional as implemented by Perdew, Burke, and Ernzerhof (PBE-GGA)\cite{perdew96} and full-relativistic norm-conserving pseudopotentials~\cite{oncv}. The self-consistent calculations are found optimally converged with basis set energy cutoff of 35 Ry and 16 $\times$ 16 $\times$ 1 Monkorst pack \textit{k}-mesh grid. We constructed maximally localized Wannier functions (MLWF) \cite{mlwf1,mlwf2} using WANNIER90 package\cite{wannier90}, projected from both Mo-\textit{d} and S-\textit{p} orbitals. Shift current properties were computed using Wannier interpolation scheme \cite{sc18} with 500 $\times$ 500 $\times$ 1 integration \textit{k}-grid tested for convergence.

%

\end{document}